\documentclass[reprint,amsmath,amssymb,aps,prl]{revtex4-1}

\usepackage{mathptmx}

\usepackage{amsmath}
\usepackage{amsthm}
\usepackage{bm}
\usepackage{bbm}

\usepackage{graphicx}
\usepackage[section]{placeins}
\usepackage[caption=false]{subfig}
\usepackage{algpseudocode}
\usepackage{algorithm}
\usepackage{booktabs}
\usepackage{multirow}

\usepackage[per=reciprocal,obeyfamily,obeybold,obeyitalic]{siunitx} 

\usepackage{ragged2e}

\usepackage{xspace}

\usepackage{relsize}

\usepackage{rotating}
\usepackage{floatpag}

\usepackage[usenames,dvipsnames]{color}
\definecolor{grey}{rgb}{0.5,0.5,0.5}
\makeatletter \newcommand \listoftodos{\section*{Todo list} \@starttoc{tdo}}
\newcommand\l@todo[2]
 {\par\noindent \textbf{#2}:  #1\par\vspace{0.2cm}} \makeatother

\usepackage{braket}
\usepackage[flushleft]{threeparttable}

\usepackage[colorlinks=false, pdfborder={0 0 0}]{hyperref}
\usepackage{cleveref}
\crefname{thm}{thm.}{thms}
\Crefname{thm}{Theorem}{Theorems}

\newcommand{\dd}{\mathrm{d}}

\newcommand{\trans}{\mathsf{T}}

\newcommand{\pderiv}[2]{\frac{\partial\hspace{-0.1em} #1}{\partial\hspace{-0.1em} #2}}

\DeclareMathOperator{\tr}{tr}

\newcommand{\SIESTA}{\textsc{Siesta}\xspace}

\newcommand{\EOS}{\textsc{eos}\xspace}

\begin{document}

\title{Dynamical continuum simulation of condensed matter from first-principles}

\author{Oliver Strickson}
\altaffiliation{Current address: The Alan Turing Institute, British Library, 
96 Euston Road, London NW1 2DB}
\email{ostrickson@turing.ac.uk}
\affiliation{Cavendish Laboratory, University of Cambridge, 
J. J. Thomson Avenue, Cambridge CB3 0HE, United Kingdom}
\author{Nikos Nikiforakis}
\affiliation{Cavendish Laboratory, University of Cambridge, 
J. J. Thomson Avenue, Cambridge CB3 0HE, United Kingdom} 
\author{Emilio Artacho}
\affiliation{Cavendish Laboratory, University of Cambridge, 
J. J. Thomson Avenue, Cambridge CB3 0HE, United Kingdom}

\begin{abstract}
  Macroscale continuum mechanics simulations rely on material 
properties stemming from the microscale, which are normally 
described using phenomenological equations of state (EOS). 
  A method is proposed for the automatic generation of first-principles 
unconstrained EOSs using a Gaussian process on a set of ab initio 
molecular dynamics simulations, thereby closing the continuum equations. 
  We illustrate it on a hyperelasticity simulation
of bulk silicon using density-functional theory (DFT), following the
dynamics of shock waves after a cylindrical region is instantaneously 
heated.
\end{abstract}

\maketitle


  Continuum mechanics simulations are of great importance for
the simulation of macroscopic condensed matter, from 
flows in  the oil and gas industry \cite{koblitz2018,sverdrup2018} 
or the modelling of high strain-rate structural deformation \cite{michael2018}, 
to shock waves and detonation in condensed-phase 
media \cite{schoch2013,schoch2013multi,schoch2013eulerian}. 
   The phenomena these simulations describe have their 
origin in the interactions and dynamics of electrons 
and nuclei at a scale much smaller than the one of the continuum 
simulation. 
  Continuum mechanical theories disregard the constituent particles,  
however, using phenomenological equations of state (EOS) 
\cite{eliezer2002}, to which a substantial research effort is dedicated 
(see, e.g. \cite{menikoff2009,heuze2012,menikoff2015,wilkinson2017,
weck2018,rakhel2018}).
  At the atomic scale, electronic structure methods allow 
the accurate and predictive evaluation of the EOS of 
condensed matter from first-principles 
\cite{swift2001first,correa2018}.
  Indeed, this was one of its earliest applications, 
predicting cold pressure-volume curves for isotropic simple 
materials \cite{cohen1982}. 
  An equation of state is generally much more complex, 
including all deformations beyond linear, and temperature, 
for a total of seven dimensions in the case of a solid.

  Here we describe a procedure for computing an EOS from first 
principles in a general way, that neither depends on the material, 
imposes any functional form,  nor requires parameter fitting.
   This contrasts with the use of traditional EOS's,
such as Mie-Gr\"uneisen \cite{eliezer2002,hiermaier2008}.
   The generality and accuracy of the former is only limited by
those of the underlying first-principles theory,
although the latter keeps an advantage in efficiency.

\begin{figure}[t]
\centering
\includegraphics[width=0.33\textwidth]{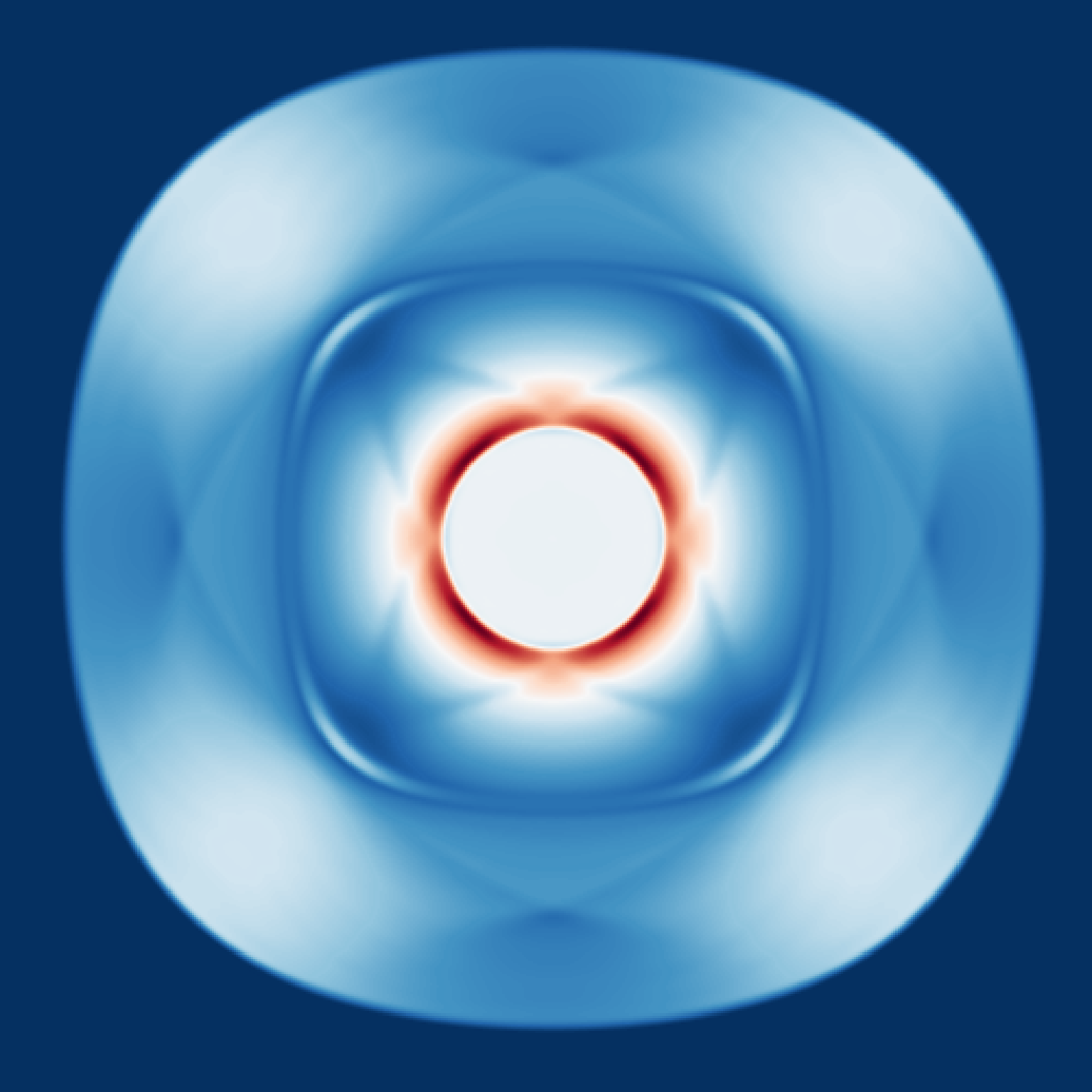}
\makeatletter
\renewcommand{\@caption@fignum@sep}{. }
\makeatother
\caption{Deviatoric stress for a shock wave
  in bulk silicon at  zero pressure and \SI{40}{K} from the sudden 
  heating of a macroscopic cylindrical region of radius $R$ (white circle).
  Snapshot taken at time $t=R/\alpha$ after the heating onset,
  with $\alpha=10^4$ m s$^{-1}$.  
  The square domain is 10 $R$ wide.
  The lowest value in the figure (0 MPa)
  is indicated in dark blue, the highest (285 MPa)
  with dark red. In the inner cylinder it is 136 MPa.
  The EOS for silicon is obtained as described here.} 
\label{fig:hot-cyl-1}
\end{figure}

\begin{figure*}[htb]
\centering 
\includegraphics{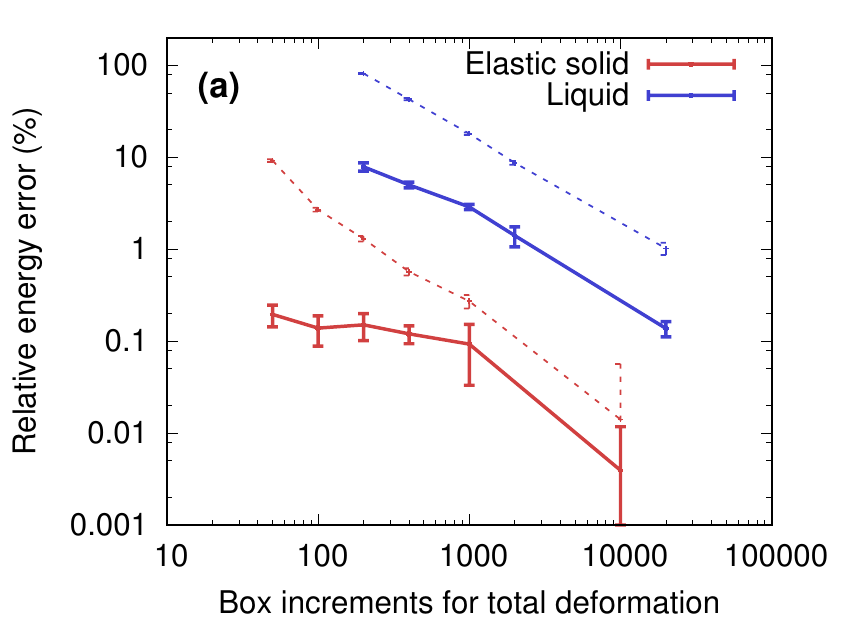}
\includegraphics{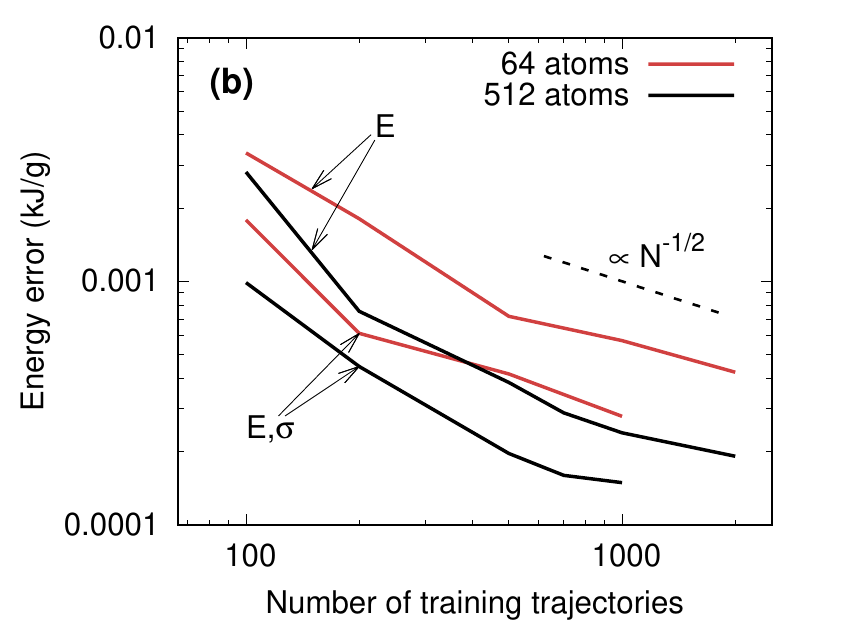} 
\makeatletter
\makeatother
\caption{Sources of error.  \textbf{(a)} Convergence in energy 
for isentropic trajectories, integrating the work done by the
deforming box (solid lines) and monitoring the internal energy
(dotted lines). 
  Red (blue) lines are for crystalline (liquid) silicon deformed to a 
uniaxial strain of 0.9 (0.8).
  Points and error bars indicate ensemble averages sampling 
equivalent trajectories. 
  \textbf{(b)} Reconstruction error from the Gaussian
process versus number of isentropic deformation trajectories, 
training from energy alone (`E') or energy and stress (`E,$\sigma$'). 
}
\label{fig:energy-difference}
\label{gp-errors}
\end{figure*}

  Our EOS is constructed using a machine-learning 
Gaussian process that probes the relevant space by a series of
ab-initio molecular-dynamics (AIMD) simulations, which inform the
continuum simulations.
  We illustrate the method with DFT for hyperelastic 
solids, but the procedure is applicable much more generally in 
systems and underlying theory.
  \Cref{fig:hot-cyl-1} shows results 
for shock-waves emanating into bulk
silicon as described by DFT using the method in this
Letter, at the level of nonlinear, anisotropic hyperelasticity.  
  The initial condition was a cylindrical inclusion of a 
high temperature region, described in more detail below.

  For the continuum simulations, the numerical solution of a 
system of conservation laws is obtained 
in the Eulerian frame \cite{Godunov1972}.
  A general deformation is represented by the
deformation gradient tensor,
$\bm{F}=\partial \bm{x} / \partial \bm{X}$,
where $\bm{X}$ is a material point in the undeformed configuration,
and $\bm{x}$ is its displaced position.  The system evolves in time
according to \cite{Godunov1972}
\begin{alignat}{3}
  \label{eq:def}&(\rho F_{ij})_t &&+\, (\rho F_{ij} u_k - \rho F_{kj} u_i)_k &&= 0 \\
  \label{eq:mom}&(\rho u_i)_t &&+\, (\rho u_i u_j - \sigma_{ik})_k &&= 0 \\
  \label{eq:ene}&(\rho E)_t &&+\, (\rho u_k E - u_i\sigma_{ik})_k &&= 0,
\end{alignat}
$\rho$ being the mass density, ${\bm{u}}$ the velocity field, 
$E$ the internal energy, and $\bm{\sigma}$ the Cauchy stress,
along with the initial constraint that
\begin{equation} 
  \nabla \times \bm{F}_i = 0.
\end{equation}
  An EOS closes the system giving the internal energy 
for any deformation. 
  For the definition of a symmetric strain tensor (excluding rotations)
we use the right Cauchy--Green tensor, 
\begin{equation}
\bm{G} = \bm{F}^{\trans}\bm{F}.
\end{equation}
The Cauchy stress $\bm{\sigma}$ is given in terms of $\bm{G}$ as
\begin{equation}
\bm{\sigma} = 2\rho F_{il}\left(\pderiv{E}{G_{lm}}\right)_S F_{jm},
\end{equation}
where the derivative for stress is at constant entropy: this would
suggest as most convenient a form of equation of state of
$E(\bm{G},S)$. 
  There is, however, no need for obtaining entropy values,
which would be expensive in an AIMD setting.
  The calculation of strain derivatives of the energy for
(different values of) constant entropy is what is needed.
  Instead, we can define the reference tag $E_0$ as the internal energy
(kinetic energy of nuclei plus total electronic energy) 
of the material if it were adiabatically brought to the reference 
configuration from the deformed configuration. 
  As defined, the mapping between $E_0$ and entropy does not depend 
on deformation, and, therefore, $E_0$ can be used to label the isentropes.
  The equation of state is expressed then as $E(\bm{G},E_0)$, and one
obtains $(\partial E/\partial \bm{G})_S$ from an entropy preserving (thermally 
isolated) quasistatic AIMD simulation for a given $E_0$.
That is, we follow isentropes but do not need to know the value of $S$.

  The numerical solution to \cref{eq:def,eq:mom,eq:ene} has been extensively 
discussed \cite{Godunov1972,Plohr1988,Miller2001,Barton2009}.
  A finite volume formulation is used here, with 
fluxes from the \textsc{force} scheme \cite[ch.7]{Toro2013} using the
\textsc{mpweno-5} reconstruction \cite{Balsara2000}.
  The finite-volume formulation in the Eulerian frame  
allow the capture of correct weak solutions (shock waves).
   The particular formulation of non-linear elasticity and the method of 
solution used here illustrate the use of a multidimensional EOS, 
but the same ideas can be readily ported to other situations.


  The EOS used for this simulation was obtained from 
first-principles simulations of silicon based on DFT.  
  AIMD simulations were performed with the \SIESTA method and 
implementation of DFT \cite{Soler2002}, using the PBE \citet{Perdew1996}
exchange-correlation functional.
  The basis functions for the valence electrons, and the pseudopotential
for the Si core electrons are the same as described in \cite{Strickson2015}.
  The mesh used for integrals in real space was well converged with a grid
cutoff of \SI{100}{\,Ry}.
  A $2^3$ grid of {\bf k}-points was
used on the 64 atom simulations, to give an effective cutoff length of
\SI{11}{\,\angstrom} \cite{Moreno1992}.
  Thermal electronic contributions are expected 
to be small at the temperatures considered, and 
an electronic temperature of \SI{300}{\,K} was used throughout.

Verlet integration (modified as described below) was used to follow an
isentrope, with a timestep of \SI{1}{\,fs} and forces from DFT.  480 
separate deformations of a 64-atom box were performed, 
with AIMD runs of \SI{2}{ps}.  For each deformation, 
an initial configuration was obtained by
equilibrating the system using the Tersoff empirical potential \cite{Tersoff1986}
on the intended undeformed state, before switching to DFT forces and 
continuing the integration.  The DFT dynamics was further integrated 
for \SI{250}{fs} before starting the deformation, and for \SI{250}{fs} after 
finishing it, in order to obtain averaged final quantities.


  States along an isentrope are extracted
directly with molecular dynamics in a slowly deforming box. 
  An alternative procedure is suggested in Ref.~\citet{Chentsov2012}, 
who use (for a liquid) a sampling in density and
temperature before solving an ODE to find the internal energy as a
function of density and entropy.
  In our direct procedure, AIMD gives an
isentropically deformed state to a given target deformation, starting
from an undeformed reference state at a given (randomly-sampled)
$E_0$ value, obtained by equilibrating to a given temperature.
  The box is steadily deformed
by slowly varying the box vectors in a linear process from the 
undeformed to the target deformed state.
  The entropy change due to varying them is made arbitrarily 
small by decreasing their rate of variation, since a slowly varied parameter
of a Hamiltonian preserves the entropy to first-order in the rate of
variation of the parameter (see e.g.\ \cite{LandauLifshitz5}).
  To demonstrate that we can follow an isentrope numerically, we
show that the process is adiabatic and reversible. 
 That is
\begin{equation}
dE \rightarrow \frac{1}{2} V \tr (\bm{\sigma}^\trans\dd{\bm{G}}) 
\end{equation}
as the time derivative of the deformation vanishes,
and additionally, that if the process is reversed, the energy
difference between the initial and final states tends to zero.  It is
important to note that the quantities involved in this expression are
equilibrium ensemble averages.

\Cref{fig:energy-difference} (a) shows both the difference in total energy
from this process and the integrated work.  For an isentropic process,
both should be zero, and any difference is systematic error introduced
by the process.  Two cases are illustrated: a uniaxial elastic
compression of 0.9 relative volume (representative of the deformations
we consider), and an uniaxial compression of a liquid, to 0.8 relative
volume.  The latter case is more challenging since there is additional
time for relaxation of the fluid to a hydrostatic stress
configuration.
  We can therefore apply deformations to stresses of tens of \si{GPa}
over $\sim$\SI{1}{ps} on 64 atom cells and achieve relative errors in
the total energy related to the strain of around 1\%, and error in the strain
energy computed by integrating the work of 0.2\%.

\begin{figure}[t] 
\centering
\includegraphics{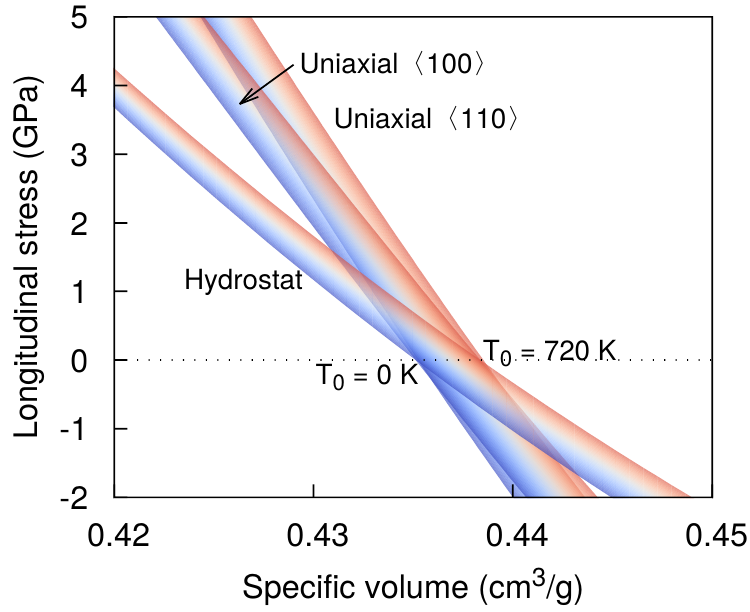} 
\makeatletter
\makeatother
\caption{Compressive isentropes from the
  first-principles equation of state for silicon.  Each compression
  (uniaxial and hydrostatic) is shown for a range of isentropes,
  coloured by their initial temperature.}
\label{fig:isentropes}
\end{figure}


\Cref{fig:isentropes} shows slices through the equation of state
for silicon used to produce \cref{fig:hot-cyl-1} 
From AIMD we have a discrete sampling of the energy surface.
The points must be interpolated to evaluate the energy of a
particular arbitrary deformation at a given temperature.  
A suitable interpolation method and a procedure for choosing 
the sampling points are crucial components of this scheme.
We use Gaussian process regression for the interpolation
\cite{Mackay2003,Rasmussen2006} for several reasons.
  First, its ability to handle multi-dimensional data. 
  Second, the fact that (with a suitable covariance function) the
interpolated function is smooth: we require the interpolant to have
continuous second derivatives, since these appear in expressions for
the wave speeds. We thereby avoid unphysical wave splitting.
  Third, it can incorporate derivate observations (e.g. from 
pressure) into the learning process, and predict 
derivatives of the interpolated function (and therefore pressures).

The Gaussian process prediction takes the form
\begin{equation}
\hat{t}' = \bm{k}^\trans \bm{C}^{-1}\bm{t},
\end{equation}
where $\bm{t}$ is the vector of observed values (total energies and
their derivatives with respect to $\bm{G}$) and $\bm{C}$
is the covariance matrix, computed from the training data as
\begin{equation}
  C_{ij} = C(\bm{x}^{(i)}, \bm{x}^{(j)}) + \nu^2\delta_{ij},
\end{equation}
where $\bm{x} = (\bm{G}, E_0)$ is the vector of inputs, and
with the squared exponential covariance function 
between energy observations, 
\begin{equation}
\label{eq:covar}
  C(\bm{x}^{(1)},\bm{x}^{(2)}) = \zeta^2 \exp \left(
    -\frac{1}{2}\sum_{i,j} \frac{(G^{(1)}_{ij} -
      G^{(2)}_{ij})^2}{r_{G_{ij}}^2} - \frac{(E_0^{(1)} - E_0^{(2)})^2}{r^2_{E_0}} \right).
\end{equation}
  The vector
$\bm{k}$ contains the covariances of the input to predict,
$\bm{x}^*$, with each of the observations; that is,
\begin{equation}
  k_i = C(\bm{x}^{(i)},\bm{x}^*).
\end{equation}
Covariances between value and derivative observations, and between two
derivative observations, are the corresponding derivatives of the
$C(\bm{x}^{(1)},\bm{x}^{(2)})$ function.

\begin{figure}
\centering
\includegraphics[width=0.93\columnwidth]{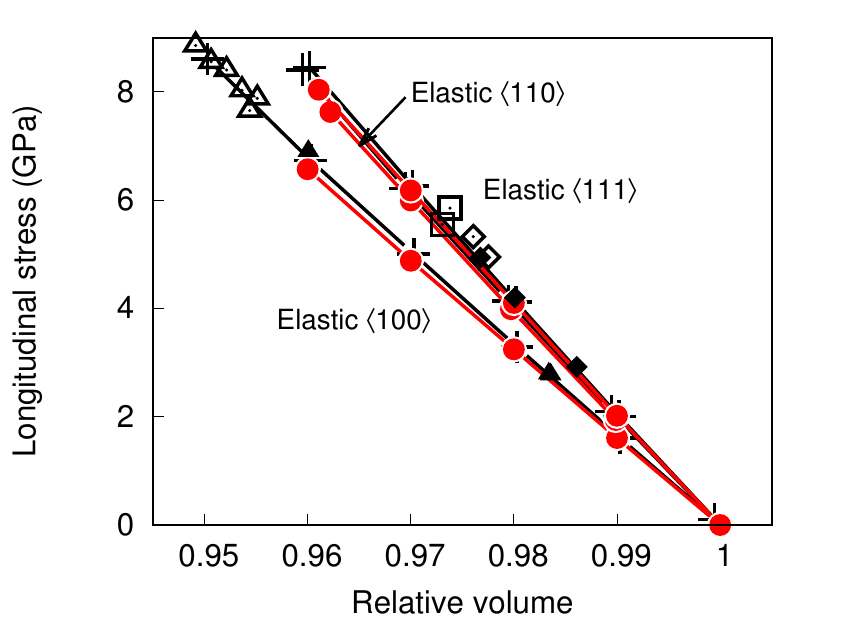}
\includegraphics[width=0.93\columnwidth]{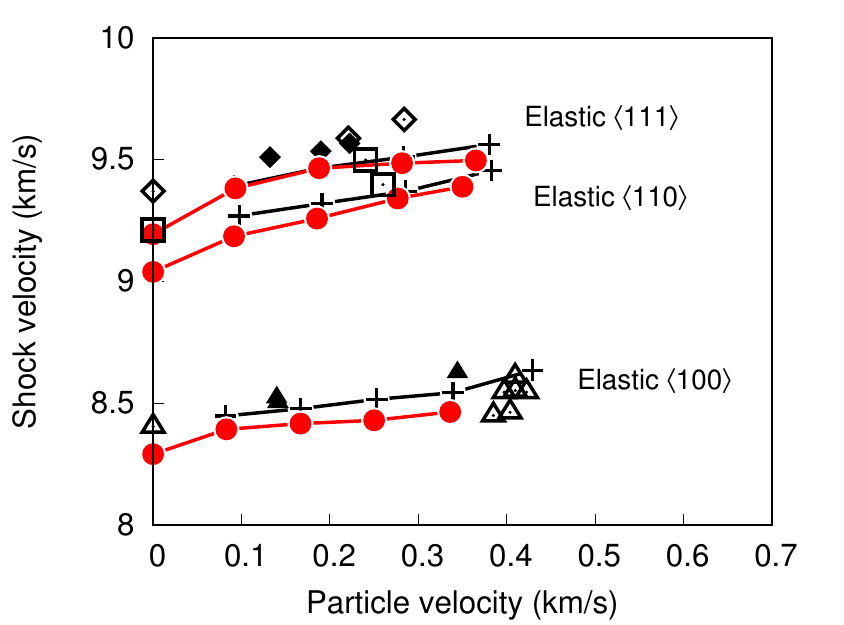}
\makeatletter
\makeatother
\caption{Validation for shock waves. Hugoniot 
states for silicon across a shock wave from an
initially uncompressed state at \SI{300}{K}. 
  Red lines with solid circles: results of this work,
black lines (+) from the direct atomistic computation of
the Hugoniot locus for the same DFT silicon
\cite{Strickson2015}. 
    Experimental results are shown for refs.\ \cite{Goto1982} 
(hollow symbols) and \cite{Turneaure2007x} (solid symbols): 
triangles represent shocks along $\langle 100 \rangle$, 
squares for $\langle 110 \rangle$, diamonds for $\langle 111 \rangle$.}
\label{fig:shock-validation}
\end{figure}

The interpretation of the hyperparameters in Eq.~\ref{eq:covar}
is as follows: $\zeta$ sets the scale of
the inferred function, $\nu$ represents position-independent Gaussian
noise in the outcomes that is independent of the inputs, and
$r_{G_{ij}}$ is the length scale over which the function varies with $G_{ij}$. 
Larger values indicate less rapid dependence on the input.
Separate noise hyperparameters are used for
value and derivative observations.

The sampling is performed by choosing $\bm{G}$ uniformly at random
over a problem-specific domain of interest, before converting it to
a deformation gradient $\bm{F}$ (by a Cholesky decomposition), and
thence to a target lattice $\bm{F}\bm{L}$, where $\bm{L}$ is the matrix 
whose columns are the lattice vectors. 
For larger dimensionality other samplings may be more suitable
\cite{Simpson2001}.

\begin{figure*}[t]
\centering
\includegraphics[width=0.33\textwidth]{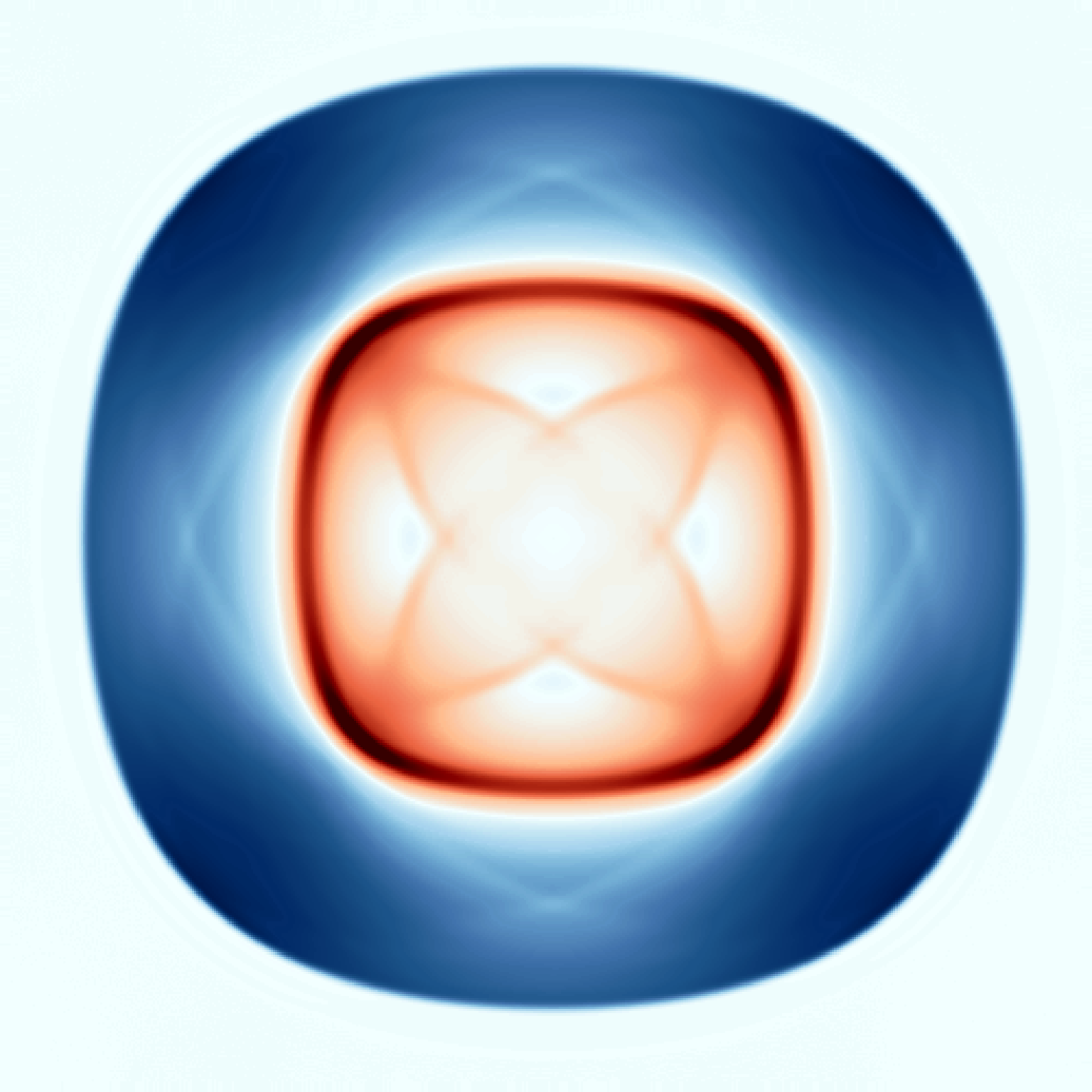}
\includegraphics[width=0.33\textwidth]{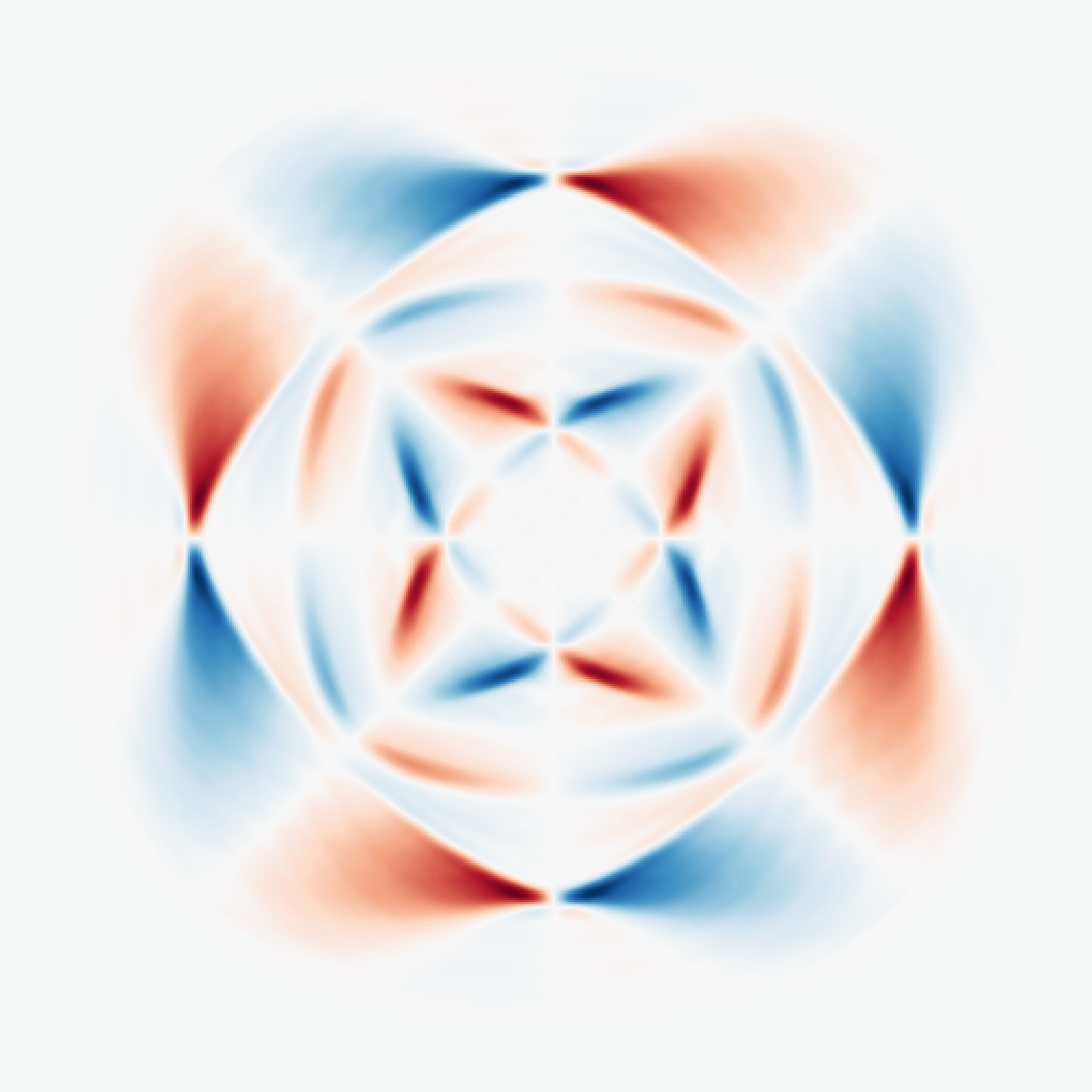}
\makeatletter
\makeatother
\caption{Hot cylinder: Radial material velocity (left)
  and transverse material velocity (right), for a shock wave 
  in bulk silicon from sudden cylindrical heating as in \cref{fig:hot-cyl-1}.  
  The lowest values of the radial ($-$12.5 m/s) and 
  transverse material velocities ($-$1.5 m/s) are indicated in dark red.
  The highest respective values (8.5 m/s and 1.5 m/s) 
  are indicated in dark blue.
  The deviatoric stress is shown in \cref{fig:hot-cyl-1}.}
\label{fig:hot-cyl-2}
\end{figure*}

The sampling domain can be chosen generously to include the range over
which the deviatoric part of the strain is expected to be less than or
equal to the yield criterion, according to, for example, a continuum
plasticity model, and with the isotropic part of the strain less than
some bound.  For the EOS given here, we
sample each component independently uniformly over the range 
$[0.9,1.1]$ for the diagonal components and $[-0.3,0.3]$ for the
off-diagonal ones.  The internal energy of the undeformed state 
$E_0$ is sampled by varying the initial $T_i \in [0,800]$ K.
  Since $E_0$ is the dominant contribution to the internal energy,
the fitting is improved by defining $E'$
\begin{equation}
E'(\bm{G},E_0) = E(\bm{G},E_0) - E_0
\end{equation} 
as the quantity to interpolate.

The error from the reconstruction is shown in \cref{gp-errors}(b), from
an equation of state computed from molecular-dynamics trajectories
from an empirical potential, allowing larger sampling.  For
the databases where gradient information is used, all six components
of the gradient are included for one-sixth of the points in the
database.  The figure shows that this is always beneficial, 
but much more so for small databases, where it can reduce
the error by a factor of four.
In addition, if symmetry is exploited, 
the sampling efficiency is increased by a factor that depends 
on the crystal system (8 for cubic).


  Validation of the multi-scale method proposed in this work is 
provided by the comparison shown in \cref{fig:shock-validation} 
for properties of silicon shocked with a flat two-dimensional perturbation.
  The properties obtained here are compared with experimental
results as well as with independent simulations results for the same DFT silicon 
obtained from an ab initio Hugoniot calculation \cite{Strickson2015}.
  The agreement is highly satisfactory. 
  In addition, a full, explicit first-principles shock wave has been simulated 
using AIMD with the same DFT as used here for the EOS. 
  A 2$\times$3$\times$20 supercell with 960 Si atoms was
pushed with a piston along the (001) direction with a velocity of 360 m/s.
  The velocity of the ensuing shock wave calculated with the method described 
in this Letter was  2\% higher than the one obtained from 
the explicit AIMD simulation, offering again a satisfactory validation of
the method.

  It should be remembered, however, that the method described here
is of a much more general applicability than flat shock waves,
while the method in Ref.~\cite{Strickson2015}
is only valid for such shocks, making explicit use of the Hugoniot relations.
Figures~\ref{fig:hot-cyl-1} and \ref{fig:hot-cyl-2} illustrate a 
much more general case that cannot be simulated otherwise, namely,
for a shock wave generated in bulk silicon from the sudden heating 
to \SI{600}{K} of a cylinder of radius $R$ in zero-pressure \SI{40}{K} bulk silicon.
  The figures show the behavior of the deviatoric stress (\Cref{fig:hot-cyl-1}),
the radial material velocity [\Cref{fig:hot-cyl-2} (a)] and the transverse
material velocity [\Cref{fig:hot-cyl-2} (b)] at a time $t=R/\alpha$ after the initial 
shock, with $\alpha=10^4$ m s$^{-1}$.
  The initial cylindrical shock is deformed into the displayed shapes 
due to the anisotropy of the material.
  There is a scale invariance in the continuum equations that allows
$R$ to be macroscopic, which is out of reach for purely atomistic 
simulations.


In summary, a two-scale method has been demonstrated for 
the generation of EOSs for macroscopic continuum simulations 
of condensed matter based on first-principles 
molecular dynamics. 
  The AIMD simulations are performed on a sample of points selected by a 
machine-learning Gaussian process in the space of parameters,
for the required EOS to be effectively interpolated to any other point, 
as requested by the continuum mechanics simulation.
  As a first step, it has been illustrated on complex hyperelastic 
shock waves in bulk silicon as obtained from DFT calculations, 
for which the method has been validated.
  Condensed matter systems of other forms or in other regimes, such as liquids, 
glasses, polycrystalline solids or solids under plastic deformation, would also be
amenable to this method or extensions thereof, using continuum techniques 
(e.g. assuming yield behaviors for plastic deformation \cite{simo2000}) and 
MD simulations at larger scales \cite{chandra2018}.
  The method described in this paper brings first principles to a wide range of
continuum mechanics, including for 
materials that have not been synthesized yet. 

{\it Acknowledgments.} 
  We are grateful for the support of Dr. Alan Minchinton. 
  OS acknowledges funding from Orica Limited through grant RG63368.
  Calculations were performed on the Darwin Supercomputer of the 
University of Cambridge High Performance Computing Service, using 
Strategic Research Infrastructure Funding from the Higher Education 
Funding Council for England, and funding from the Science and 
Technology Facilities Council.



%

\end{document}